# All-carbon nanotube solar cell devices mimic photosynthesis

*Gideon Oyibo[1], Thomas Barrett[1], Sharadh Jois[1], Jeffrey Blackburn[2], Ji Ung Lee*[1]*

[1] College of Nanoscale Science and Engineering, State University of New York-Polytechnic Institute, Albany, New York 12203, USA.

[2] National Renewable Energy Laboratory, Golden, Colorado 80401, USA

KEYWORDS : carbon nanotube, photovoltaic, photosynthesis, excitons, energy transfer

**Abstract**

Photovoltaics has two main processes: Optical absorption and power conversion. In photosynthesis, the two equivalent processes are optical absorption and chemical conversion. Whereas in the latter, the two processes are carried out by distinct proteins, in conventional photovoltaic diodes, the two processes are convoluted because the optical and transport paths are the same, leading to inefficiencies. Here, we separate the site and direction of light absorption from those of power generation to show that semiconducting single-walled carbon nanotubes (s-SWCNTs) provide an artificial system that models photosynthesis in a tandem geometry. Using different s-SWCNT chiralities, we implement an energy funnel in dual-gated *p-n* diodes. This enables the capture of photons from multiple regions of the solar spectrum and the funneling of photogenerated excitons to the smallest bandgap s-SWCNT layer, where they become free carriers. As a result, we demonstrate an increase in the magnitude and spectral response of photocurrent by adding more s-SWCNT layers of different bandgaps without a corresponding deleterious increase in the dark leakage current.

**TOC Graphic**

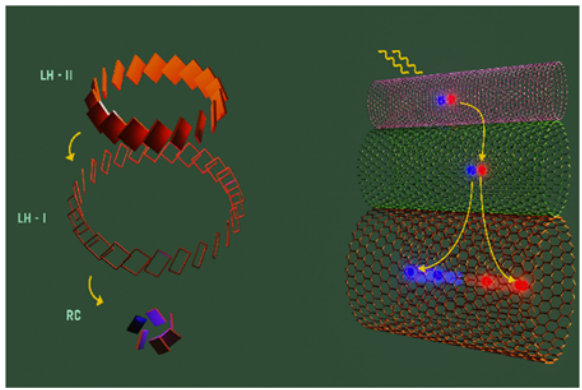

Nature has adopted an efficient way to harvest sunlight and convert it to energy in photosynthetic systems. Photosynthesis in plants, algae, and photosynthetic bacteria relies on light-harvesting complexes that capture light from multiple regions of the solar spectrum and convert it to excitons. For added efficiency, these systems have also adopted an energy hierarchy that funnels excitations from their light-harvesting complexes to their reaction center (RC), where chemical conversion occurs[1–5]. Thus, photosynthetic systems have separated the two main processes of optical absorption and chemical conversion using distinct proteins. Mimicking these two processes in artificial energy conversion systems has always been an important goal for researchers across various disciplines. While the reaction features of natural photosynthesis have been mimicked in photoelectrochemical systems[6–8], this work aims to design a photovoltaic diode that models photosynthesis by separating the site and direction of light absorption from that of energy conversion, as shown in figure 1a. This architecture allows each process to be optimized independently, unlike conventional solar cells that convolute these processes. We estimate that this design could improve the efficiency of solar cells by at least 29% over conventional single-junction solar cells (See supplementary information).

To mimic photosynthetic systems, we use s-SWCNTs of varying bandgap in a tandem design. The technique we develop can be used with a larger set of s-SWCNTs to capture light from the broad solar spectrum. Conventional tandem solar cells already accomplish this with multiple junctions but use complex fabrication techniques at a high cost. The detailed balance limit for the efficiency of tandem solar cells is predicted to reach 68% for unconcentrated light and 86% for concentrated light[9]. Although achieving such efficiencies is far beyond the current state-of-the-art[10], using a single absorber material with tunable bandgaps promises to make the fabrication easier, as we demonstrate below.

Single-walled carbon nanotubes are rolled-up sheets of graphene that can form a large variety of chiralities depending on the diameter and roll-up angle of the nanotube. They can be either metallic or semiconducting. s-SWCNTs have varying bandgaps that depend on their diameters, with each nanotube absorbing light at different optical transition energies. As a result, a combination of s-SWCNTs with diameters ranging from 0.6-1.6nm can cover the solar spectrum from the ultraviolet to the near-infrared[11]. In addition, s-SWCNTs have high carrier mobilities and are solution-processable, making them an ideal system to model photosynthesis as the active layer in solar cell devices[11–15].

While the study of the photovoltaic properties of individual nanotube p-n diodes is already an established field[16–18], they can only harvest a tiny fraction of incident photons because of their limited area and bandgap. There is therefore a need to explore the light-harvesting potential of a network of s-SWCNTs with different bandgaps in *p-n* diodes. Purification by polymer wrapping and chromatography has enabled access to specific s-SWCNT chiralities with different bandgaps[19–22]. This progress has made it possible to fabricate heterojunction solar cells using s-SWCNTs as a donor with fullerenes and other small molecules as acceptors[23–25]. Here, we fabricate solar cell devices with s-SWCNT networks as the sole absorber since the built-in field of the *p-n* junction obviates the need for acceptor materials to dissociate photogenerated excitons. We use the following s-SWCNTs with different diameters: arc discharge (~1.55nm), (7,5) (0.83nm) and (6,5) (0.76nm) [26].

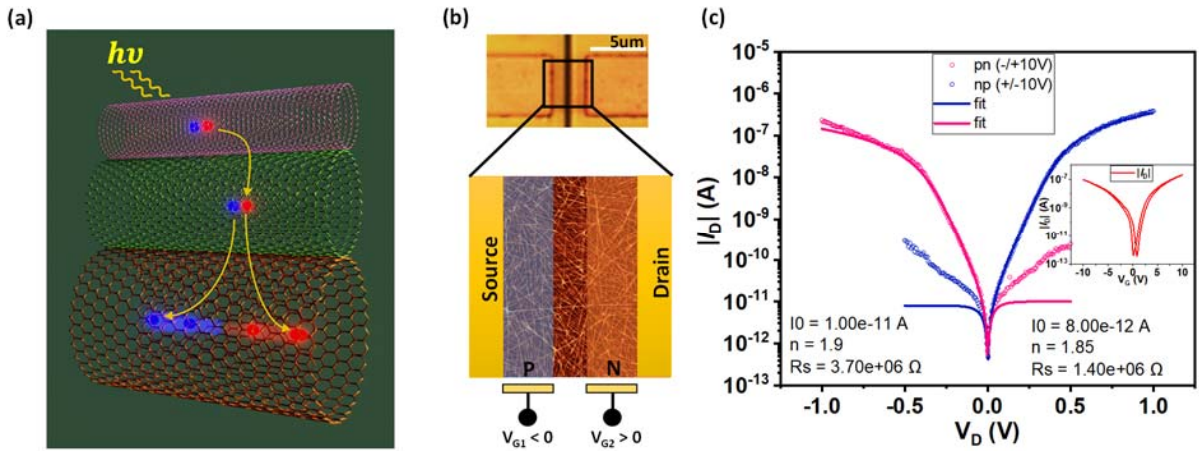

*Figure 1* **Architecture of an all-carbon nanotube tandem solar cell and diode properties**. *(**a**) A schematic of energy funnel of photoexcitations from the larger bandgap s-SWCNTs to the smallest bandgap s-SWCNT, which forms the main diode where energy conversion takes place. (**b**) Optical image of our p-n diode showing buried split gates for electrostatic doping with false-colored AFM image of arc discharge s-SWCNT network forming the main diode. (**c**) Dark I-V characteristics of the main diode for both p-n and n-p configurations showing the diode behavior comes from only the s-SWCNT channel and not the metal - semiconductor Schottky contacts. **Inset:** Transfer curve*

*of arc discharge network showing ambipolar characteristics. The back gate is swept with a fixed drain bias of -0.1 V. Split gate spacing used for the diode is 0.3 µm and length between contact electrodes is 3 µm.*

We fabricate *p-n* diodes on a network of s-SWCNTs using electrostatic gating techniques as outlined in experimental methods. We fabricate two types of network *p-n* diodes: individual chirality diodes (arc, (7,5), (6,5)) and tandem diodes ((7,5)/arc, (6,5)/arc, and (6,5)/(7,5)/arc). We fabricated dozens of devices of each type and show representative data. The different chiralities of s-SWCNTs are purified using polymer wrapping as described in methods. Using buried split gates, we form p- and n-doped regions on the s-SWCNT network by applying opposite bias, as shown in fig. 1b. To achieve ambipolar conduction, a prerequisite for obtaining good diode characteristics, we anneal and measure our devices in vacuum. The transfer curve for a large-diameter arc discharge s-SWCNT network (Fig. 1c inset), taken using the silicon back gate, demonstrates ambipolar characteristics. The current-voltage (I-V) characteristics show clear rectification behavior when biased as a *p-n* diode (Fig. 1c main panel and Fig. S1), and we fit the data to the diode equation $I_D = I_o \left( \exp\left(\frac{qV}{nKT}\right) - 1 \right)$, where *V* is applied voltage, *KT* is thermal energy, $I_o$ is the dark leakage current, and *n* is the diode ideality factor. The ideality factor *n* is 1 for ideal diodes and 2 for diodes with mid-gap states, which could be induced by the substrate and/or polymer residue from the purification and fabrication process. By fitting the forward bias region of the I-V curve, we extract $I_o$, *n*, and the contact resistance, R, which we observe at high bias. These diodes show non-ideal diode behavior with n = 2 ± 0.5.

Figure 2a-c shows the sequential steps used to fabricate the tandem s-SWCNT *p-n* diodes (bottom panels), along with associated atomic force microscopy (AFM) images of the s-SWCNT network

deposited in each step (top panels). We measure the layer thickness using AFM and estimate the number of s-SWCNT layers by dividing the AFM step height (Fig. 2d) by the diameter of the s-SWCNT[26]. Figure 2e shows the expected inverse diameter dependence of the radial breathing vibrational mode (RBM) measured by Raman spectroscopy of each isolated s-SWCNT network.

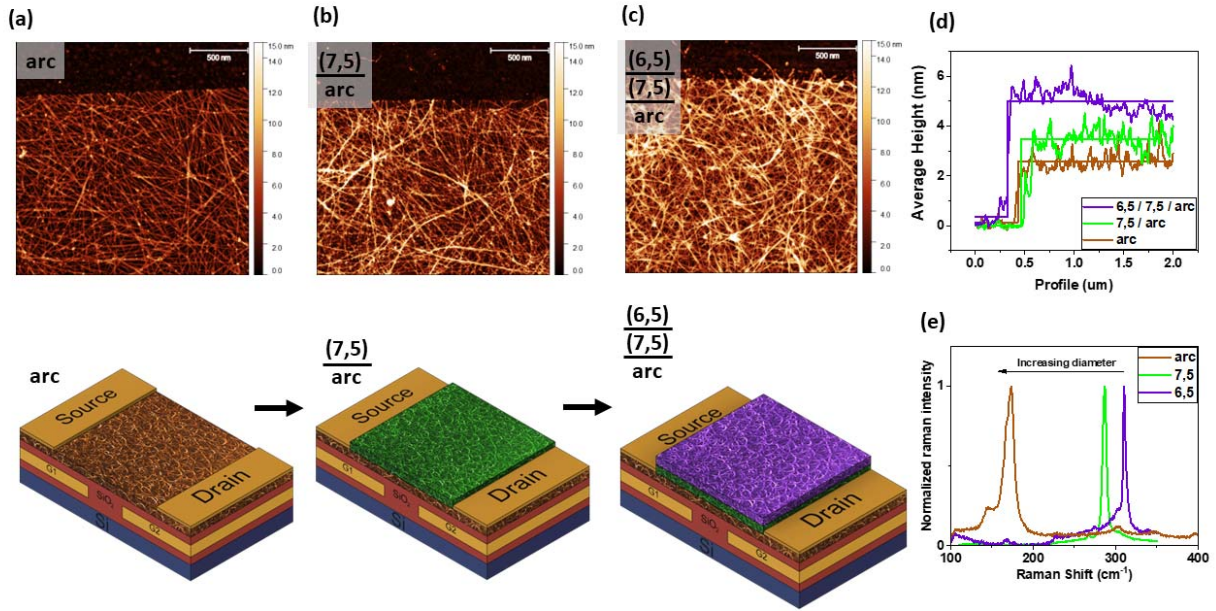

*Figure 2| **Evolution of device fabrication**. AFM images after channel etch for network of (**a**) arc, (**b**) (7,5)/arc and (**c**) (6,5)/(7,5)/arc s-SWCNTs with schematics showing the device evolution as different layers are added. (**d**) Step height of s-SWCNT networks measured by averaging 512 line scans from each AFM map. This shows that we deposited 1.5 layers of arc, a monolayer of (7,5) and 2 layers of (6,5) s-SWCNTs. (**e**) Raman spectra for arc, (7,5) and (6,5) s-SWCNT networks showing their RBM peaks.*

Figure 3a displays the photocurrent spectra from individual chirality s-SWCNT diodes. It is established that excitons (photogenerated electron-hole pairs bound to each other by their Coulomb binding energy, $E_b$) dominate the optical transitions in the visible and near-infrared range[11,27]. Photocurrent spectra in Fig. 3a reflect the dominant excitonic transitions which is consistent with the absorbance spectra for each sample (Fig. S7)[19,22]. We measured devices of different thicknesses and show that the excitonic spectral weight per s-SWCNT layer is similar across different thicknesses (Fig. S2). This correlation allows us to compare devices by normalizing to the number of s-SWCNT layers, and demonstrates that the photocurrent magnitude of s-SWCNT *p-n* diodes increases with nanotube diameter. This observation is consistent with the inverse-diameter scaling of the s-SWCNT exciton binding energy[27–29], since photogenerated excitons must be field-dissociated into free carriers to contribute to photocurrent, unless they auto-dissociate or relax to a lower energy free carrier state[11,30]. The electronic bandgap of the s-SWCNT ($E_g$) is related to the $S_{11}$ optical bandgap through the binding energy ($E_b$): $E_g = S_{11} + E_b$[11]. From the activation energy measurements of our devices, we estimate $E_g$ of each s-SWCNT network to be 0.9eV, 1.6eV and 1.8eV for arc, (7,5) and (6,5) respectively, giving binding energies ranging from ~0.25eV to ~0.56eV (Fig. S3). These values agree with theoretical predictions and experimental results from previous works that put the exciton binding energies of s-SWCNTs from hundreds of meV to 1eV [27–29,31,32]. The larger binding energies of (6,5) and (7,5) increase the probability of exciton decay before photocurrent generation from free carriers. This explains the higher photocurrent in arc s-SWCNT devices and is consistent with other results that have shown that exciton decay in s-SWCNTs increases with bandgap[33].

To construct a photosynthesis-mimicking 'energy funnel' device architecture that enhances photocurrent generation in large bandgap s-SWCNTs, we layer the s-SWCNTs according to their bandgap. Previous results have shown that downhill energy transfer (ET) between s-SWCNTs is an ultrafast process that occurs within sub-10fs timescales up to 3ps[34–38]. Since this process is faster than the rate at which photogenerated excitons recombine back to their ground state through either radiative (10 – 100 ns) or non-radiative (20 - 200 ps) mechanisms[39–45], ET should facilitate energy funneling from the smaller diameter CNTs into the arc discharge 'reaction center'. To probe the impact of ET in our energy funnel solar cells, we focus on the evolution of the diode properties as we form the (6,5)/(7,5)/arc tandem device. After forming each layer, we measure the averaged thickness, the diode I-V properties, and the photocurrent spectra. The arc s-SWCNT layer forms the main diode and mimics the 'reaction center', collecting excitons from larger bandgap s-SWCNTs and converting them to photocurrent. We sequentially deposit arc, (7,5) and (6,5) networks and achieved 1.5 layers of arc, a monolayer of (7,5) and 2 layers of (6,5) as determined from the AFM scans in fig 2d.

Figure 3b shows the photocurrent spectra taken after the sequential deposition of (7,5) and (6,5) s-SWCNT layers on the arc s-SWCNT layer. We observe a significant increase in photocurrent from (7,5) excitonic levels labeled $S_{11}^{7,5}$ and $S_{22}^{7,5}$ over the homogenous (7,5) device shown in Fig.3a. To quantify this enhancement, the photocurrent contribution from the (7,5) network is extracted from the (7,5)/arc spectra by subtracting out the arc photocurrent spectrum of the same device taken before adding the (7,5) s-SWCNT network. The extracted (7,5) spectrum is normalized to the number of 7,5 s-SWCNT layers (fig. 3c) and compared to that of the homogenous (7,5) network device. We observe a tenfold increase in the photocurrent at $S_{11}^{7,5}$ and a fivefold increase at $S_{22}^{7,5}$ transitions, supporting our hypothesis that photocurrent generation from ET to lower bandgap s-SWCNTs is more favorable than exciton dissociation in isolated (7,5) s-SWCNT devices. We observed a similar trend by comparing (6,5)/arc over (6,5) network devices (fig. 3d and fig. S4).

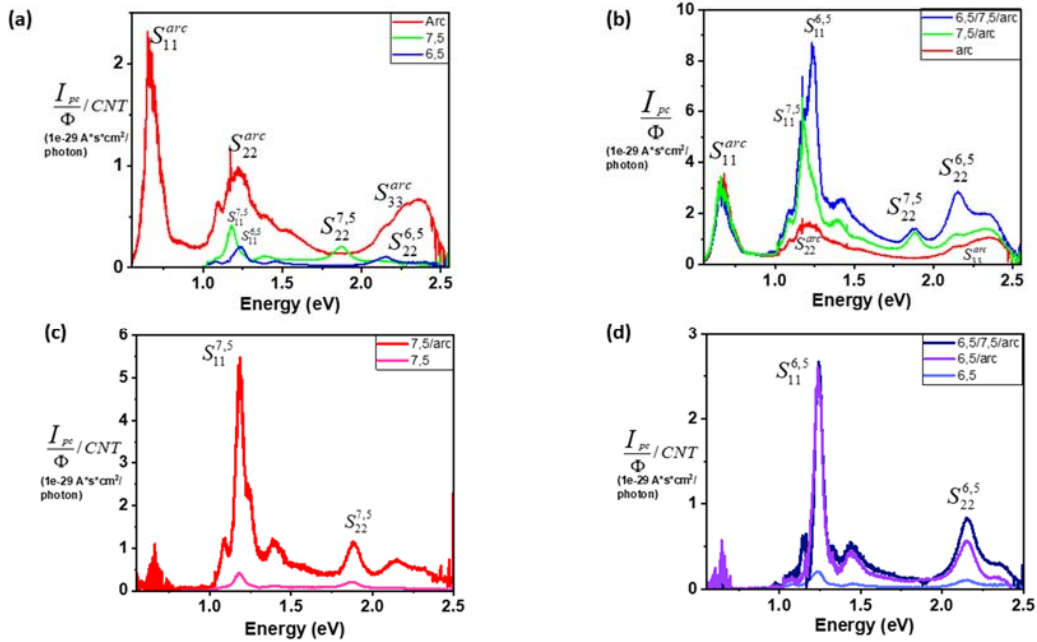

*Figure 3| **Normalized photocurrent spectra of s-SWCNT network devices** (**a**) Spectra from a homogenous network of arc discharge, (7,5) and (6,5)  (**b**) Evolution of spectrum showing an increased response from (7,5) and (6,5) after their addition to the bottom arc s-SWCNT diode. Extracted spectra for (**c**) 7,5 and (**d**) 6,5 showing a boost in photocurrent due to energy funnel.*

We observe similar levels of photocurrent enhancement from the $S_{22}$ levels of (6,5) and (7,5) s-SWCNTs (fig S5) even though they have varying degrees of overlap with the excitonic transitions of arc. Also, photocurrent enhancement from the (6,5) layer to the arc layer is equally efficient with or without an intervening (7,5) s-SWCNT layer. As shown in figure 3d, we observe similar levels of photocurrent from (6,5) s-SWCNTs in (6,5)/(7,5)/arc and (6,5)/arc devices. Surprisingly, these observations deviate from the Förster resonance energy transfer mechanism in photosynthesis which depends on both the spectral overlap and distance between donor and acceptor molecules[2].

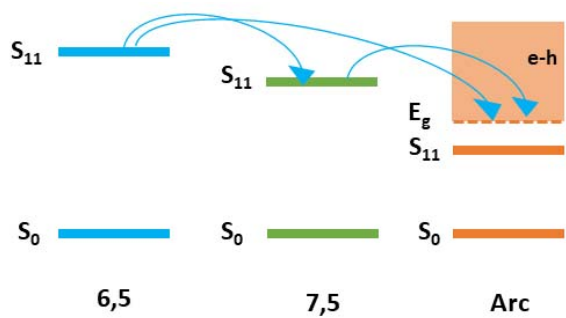

Figure 4| *Energy level diagram showing photogenerated excitons from larger bandgap s-SWCNTs relaxing to the free electron-hole continuum of arc*

Also, had the excitons from (6,5) and (7,5) relaxed to the excitonic level of arc, we would have observed similar photocurrents. Instead, the 6,5 and 7,5 $S_{11}$ photocurrent magnitudes in the (6,5)/arc and (7,5)/arc devices are at least 50% higher than the photocurrent from the arc s-SWCNTs. Thus, the primary relaxation pathway for excitons from (6,5) and (7,5) must be to the free carrier continuum of arc s-SWCNTs (See fig. S6). As such, we attribute this photocurrent enhancement to an ultrafast downhill energy transfer from (6,5) and (7,5) to the free electron-hole continuum of arc, as depicted in fig. 4. Although ET in s-SWCNTs has been studied using theoretical modeling, photoluminescence, transient absorption, and two-dimensional white-light spectroscopy[34,35,40,46], we provide strong evidence that ET leads to a boost in photocurrent generation for larger bandgap s-SWCNTs.

In photosynthetic systems, apart from an energy hierarchy to funnel photoexcitations to the RC, their major attribute is that distinct proteins carry out optical absorption and chemical conversion. Inspired by nature's design, our tandem diode also separates its two analogous processes by making the electronic path orthogonal to the optical path. In our devices, power generation is carried out by the main diode formed on the smallest bandgap s-SWCNT layer. Adding more s-SWCNTs increases the photocurrent without impacting the dark diode characteristics, which allows increased power generation. We systematically measure the diode I-V curves under dark and light conditions after the deposition of (7,5) and (6,5) layers on arc, as explained in the previous section. Figure 5a shows that the dark diode I-V characteristics remain unaltered even though we add more s-SWCNT layers to the main diode. Thus by adding more s-SWCNT layers, we increase the photocurrent magnitude and spectral response of our devices and increase the open-circuit voltage $V_{oc}$ as shown in fig 5b.

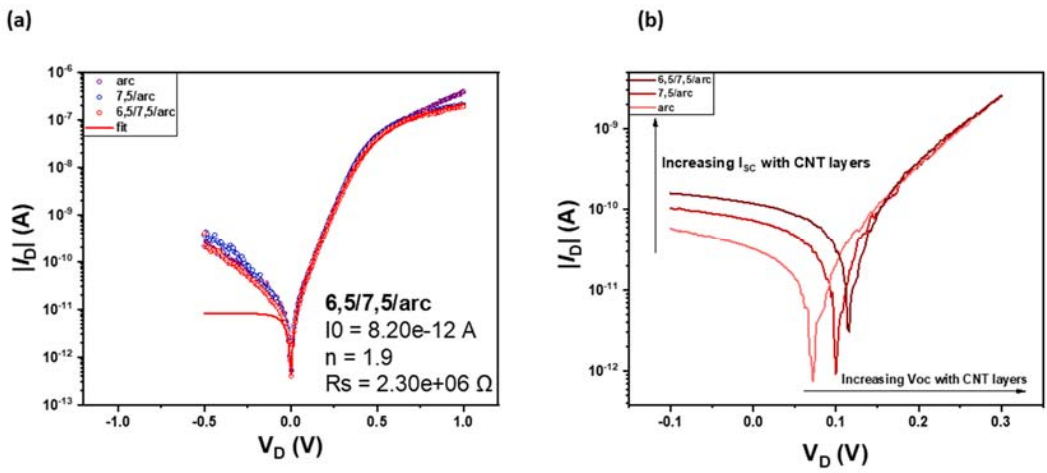

*Figure 5| **Tandem solar cell properties.** (a) dark diode I-V characteristics measured before and after the sequential addition of 7,5 and 6,5 s-SWCNT layers showing no increase in leakage current with the addition of these layers. (b) Diode I-V characteristics under illumination at 1024nm (which excites the $S_{22}$ of arc and the $S_{11}$ of 6,5 and 7,5) showing an increase in short circuit current ($I_{SC}$) and open-circuit voltage ($V_{OC}$) with the addition of s-SWCNT layers.*

In conclusion, by drawing comparisons between photosynthesis and all-carbon nanotube solar cells, we have shown clearly on a fundamental level that we can look to nature for strategies on designing more efficient tandem solar cells. If we assume photocurrent is generated at the intrinsic region of our *p-n* diodes,[16,18] we estimate our devices' peak external quantum efficiency to be 5% for devices with an intrinsic spacing of 0.3um. This efficiency is significant considering our devices consist of only a few layers of s-SWCNTs. To improve the current design, there is still the need to increase the number of s-SWCNT layers, including adding more chiralities. Even though this could potentially increase the probability of shorts and exciton quenching from metallic s-SWCNTs, outstanding recent improvements in semiconducting purity can largely preclude this issue. Also, removing the polymer used in purifying semiconducting s-SWCNTs and adopting better film deposition techniques has been shown to improve coupling between s-SWCNTs and ET[36,37]. Such improvements will increase the likelihood of greater photocurrent enhancement and higher power conversion efficiencies. Our results represent a first step towards approaching the detailed balance limit of emerging tandem solar cell technologies with semiconductors employing multiple bandgaps.

ASSOCIATED CONTENT

**Supplementary Information**.

The following are available as supporting information:

Experimental methods; Dark I-V characteristics for homogeneous 6,5 and 7,5 *p-n* diodes; normalized photocurrent dependence on thickness; excitonic transition and bandgap dependence on diameter; summary of photocurrent enhancement in (6,5)/arc and (7,5)/arc tandem devices;

UV-VIS absorbance spectra; calculations showing the advantage of using a photosynthetic design over the conventional *p-n* diode design in solar cells.

AUTHOR INFORMATION

**Corresponding Author**

*Email: leej5@sunypoly.edu

**Author Contributions**

JUL conceived and designed the study; JB purified the (6,5) and (7,5) CNTs; GO fabricated the devices, performed transport and photocurrent measurements, and carried out the data analysis; SJ and TB helped with AFM characterization, SJ and JB provided theoretical support and discussion; JUL is the principal investigator. All authors contributed to the preparation of the manuscript.

ACKNOWLEDGMENT

We would like to acknowledge the support of NREL for their help in the purification of (6,5) and (7,5) SWCNTs, CNSE metrology and AESG for on site support with tools and Sumanth Krishna for artistic contributions.